\documentclass{aip-cp}

\usepackage{graphicx}

\def\esym{$E_{sym}(\rho)$~}

\def\es0{$E_{sym}(\rho_0)$}

\def\us0{$U_{sym}(\rho_0,k_F)$~}

\def\l0{$L(\rho_0)$~}

\renewcommand{\rm}[1]{\textrm{#1}}

\begin{document}

\title{Origins And Impacts Of High-Density Symmetry Energy}

\author[aff1]{Bao-An Li}
\affil[aff1]{Department of Physics and Astronomy, Texas A$\&$M
University-Commerce, Commerce, TX 75429-3011, USA}
\eaddress{Bao-An.Li@tamuc.edu}

\maketitle

\begin{abstract}
What is nuclear symmetry energy? Why is it important? What do we know about it? Why is it so uncertain especially at high densities?  Can the total symmetry energy or its kinetic part be negative? What are the effects of three-body and/or tensor force on symmetry energy? How can we probe the density dependence of nuclear symmetry energy with terrestrial nuclear experiments? What observables of heavy-ion reactions are sensitive to the high-density behavior of nuclear symmetry energy? How does the symmetry energy affect properties of neutron stars, gravitational waves and our understanding about the nature of strong-field gravity? In this lecture, we try to answer these questions as best as we can based on some of our recent work and/or understanding of research done by others. This note summarizes the main points of the lecture.
\end{abstract}

\section{INTRODUCTION}
Answers to many challenging questions ranging from the dynamics of supernova explosions and heavy-ion collisions, the structures of neutron stars and rare isotopes to the frequency and strain amplitude of gravitational waves from spiraling neutron star binaries all depend critically on the Equation of State (EOS) of neutron-rich nucleonic matter \cite{ireview98,ibook01,dan02,Steiner05,ditoro,LCK08,Lynch09,Trau12,Tsang12,Lat13,Tesym,Chuck14,Baldo16}. The EOS in terms of the energy per nucleon $ E(\rho ,\delta )$ in nucleonic matter of isospin asymmetry $\delta=(\rho_n-\rho_p)/\rho$ and density $\rho$ is given by 
\begin{equation}\label{eos1}
E(\rho ,\delta )=E(\rho,0)+E_{\rm sym}(\rho )\cdot \delta ^{2} +\mathcal{O}(\delta^4)
\end{equation}
where $E_{\rm sym}(\rho )= \left. \frac{1}{2}\frac{\partial ^{2}E(\rho,\delta )}{\partial \delta ^{2}}\right\vert _{\delta =0}\approx E(\rho,1)-E(\rho,0)$
is the symmetry energy. Its density dependence near the saturation density of nuclear matter $\rho_0$ is normally characterized 
by its value $E_{\rm sym}(\rho_0)$ and slope $L(\rho_0)\equiv 3 \rho \frac{\partial E_{sym}(\rho)}{\partial \rho}|_{\rho_0}$. 
The Eq. \ref{eos1} is the so-called empirical parabolic law for the EOS of isospin-asymmetric nucleonic matter. Its validity within about 2 MeV has been verified by using most of the available microscopic many-body theories and phenomenological models with various realistic and/or effective interactions. However, it is worth noting that the validity of Eq. \ref{eos1} does not necessarily require the kinetic and potential part of the EOS to be individually parabolic in $\delta$ \cite{Cai-d4}. The high-density behavior of the $E_{\rm sym}(\rho)$ is currently the most uncertain part of the EOS of dense neutron-rich nucleonic matter. For example, results from using 240 Skyrme interaction parameter sets within the Hartree-Fock approach \cite{brown,Stone} and 263 parameterizations of seven different types of the Relativistic Mean Field (RMF) models \cite{RMF} were recently compared. It was found that the predicted symmetry energy spreads out rather widely \cite{Stone,RMF} especially at  supra-saturation densities. It is interesting to note that QCD-based theories were also used recently to calculate the $E_{\rm sym}(\rho)$ at supra-saturation densities with some interesting predictions \cite{Lee,Seo,Jeong}. In addition, the Auxiliary Field Diffusion Monte Carlo (AFDMC) method \cite{Gan14,Gez13}, Many-Body Perturbation Theories (MBPT) \cite{Dri14,Heb1,Heb2,Cor14}, Coupled-Cluster model \cite{Hag14},  in-medium chiral perturbation theory approaches \cite{Im-eft}, lattice chiral effective field theory (EFT) \cite{lattice} and the self-consistent Green's functions \cite{Car13a} all using two-body and three-body forces consistently derived from chiral effective field theories \cite{Epe09,Mac11,Fur13,Heb15} have been used recently to calculate the EOS and symmetry energy of neutron-rich matter. While the predictions from these calculations are remarkably consistent with each other below the saturation density $\rho_0$,  they become model dependent at supra-saturation densities where the form, strength and isospin-dependence of three-body forces, the high-momentum cut-off parameter in Chiral EFT, the isospin-dependence of short-range nucleon-nucleon correlations (SRC) and higher-order terms in both the chiral expansion of nuclear interactions and many-body perturbation theories are important but not fully understood yet. For example, the high-density behavior of $E_{\rm sym}(\rho)$ was shown to 
depend sensitively on the high-momentum cut-off parameter used in the  chiral EFT and the strength of the SRC \cite{sam14}.

The significance of nuclear symmetry energy has long been recognized by both nuclear physics and astrophysics communities. In fact, to determine the EOS of neutron-rich matter and the corresponding $E_{\rm sym}(\rho)$ is a major scientific thrust of most radioactive beam facilities around the world \cite{Ftc}. In addition, the PREX and  CREX experiments \cite{PREX-Chuck} measuring precisely the sizes of neutron-skins in $^{208}$Pb and $^{48}$Ca using parity violating electron scatterings have also been focusing on constraining the \esym \cite{PREX,Farrooh} and its implication for the radii of neutron stars \cite{Chuck-jor}. Moreover, the Neutron star Interior Composition ExploreR (NICER) \cite{nicer}will focus on investigating the EOS of super-dense neutron-rich matter in neutron stars whose many properties depend strongly on the \esym \cite{Lat13,ste14,Newton14}. There is also a strong and direct connection between the high-density symmetry energy and various features of gravitational waves from rotations or oscillations of deformed pulsars or spiraling neutron star binaries, see, e.g. ref. \cite{Far-13b,Plamen16} for a recent review. Furthermore, a thorough understanding of properties of massive neutron stars requires reliable knowledge of both the EOS of super-dense neutron-rich matter and the strong-field gravity theories simultaneously \cite{Ded03,Psa08}. Variations of the \esym can lead to large changes in the binding energy of neutron stars \cite{Newton09,He15}. There are indeed many credible theories for strong-field gravity besides Einstein's General Relativity (GR). The testing of modified gravity theories using massive neutron stars requires a reliable knowledge of the high-density \esym \cite{Wen09,Wlin14,Jiang15}.
\vspace{-0.5cm}
\begin{figure}[htb]
\includegraphics[width=5.cm,height=5.5cm]{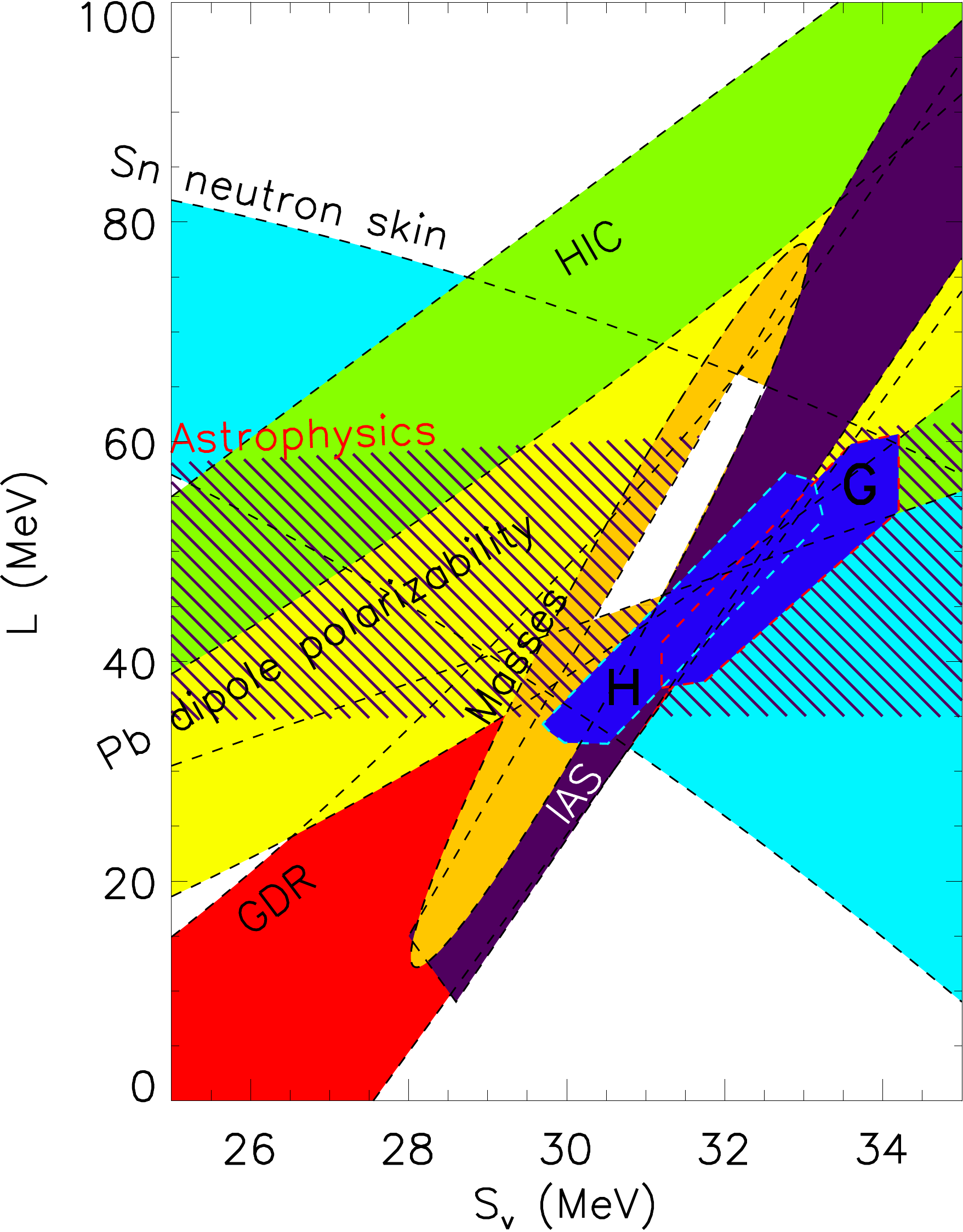}
\includegraphics[width=10.5cm,height=6.cm]{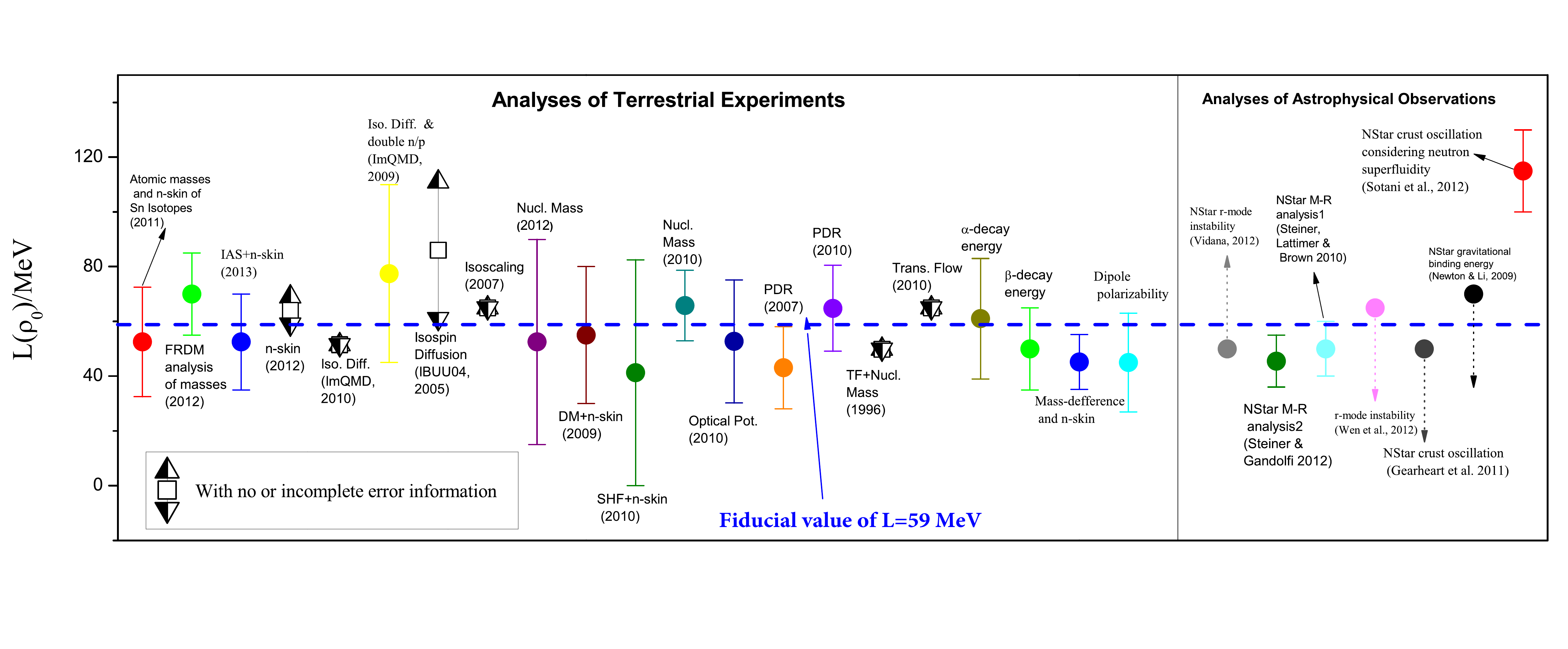}
\caption{{\protect Left: Correlation between the slope L and magnitude $S_v$ of symmetry energy at $\rho_0$ from six typical analyses in nuclear physics and astrophysics \cite{ste14}.
Right: Central values of \l0 from 28 model analyses of terrestrial nuclear experiments and astrophysical observations in the literature. Taken from ref. \cite{Li-han}.
}}\label{Jim}
\end{figure}
\section{ SYMMETRY ENERGY CONSTRAINED NEAR THE  SATURATION DENSITY}
Thanks to the great efforts made by many people in both nuclear physics and astrophysics,
significant progresses have been made recently in constraining the $E_{\rm sym}(\rho)$ around and below $\rho_0$.
Many studies have extracted the magnitude $E_{\rm sym}(\rho_0)$ and the slope parameter \l0 at $\rho_0$. Of course, they are correlated.
Shown in the left window of Fig.\ \ref{Jim} are selected six examples of recent analyses \cite{ste14}. They overlap in the area around $E_{sym}(\rho_0)\approx31$ MeV and \l0$\approx 55$ MeV.
Considering more broadly the efforts made in this field, while the $E_{sym}(\rho_0)$ is relatively better determined to be around $31.6\pm 2.66$ MeV the \l0 scatters in a rather large range and not all analyses considered the correlation between the $E_{sym}(\rho_0)$ and \l0. For example, shown in the right window of Fig. \ref{Jim} are the \l0 values scattered around a fiducial value of $59\pm 16$ MeV from 28 recent analyses \cite{Li-han}.  Overall, the consistency of the central values of \l0 from these vastly different analyses of various experiments is remarkable. 
Interestingly, the central values of $E_{\rm sym}(\rho_0)$ and \l0 extracted from the 28 analyses satisfy approximately the empirical relation $L(\rho_0) \approx 2E_{\rm sym}(\rho_0)$ which becomes exact in an interacting Fermi gas model when both the kinetic and potential symmetry energies vary with density according to $(\rho/\rho_0)^{2/3}$ \cite{LiG15}. Analyses of the very recent experiments by the ASY-EOS Collaboration found a potential symmetry energy very close to the $(\rho/\rho_0)^{2/3}$ form below about $2\rho_0$ \cite{ASY-EOS}. Naturally, all analyses are based on some models and often different approaches are used in analyzing the same data or observations.  While some of the reported constraints provide both the upper and lower limits or the standard deviation, some do not provide any information about the associated uncertainties but
only the mean values of \es0 and \l0. Moreover, the majority of the astrophysical analyses only provide the upper
or lower limits for \l0.  It is therefore currently hard to reach a community consensus about the precise value of \l0 with a quantified uncertainty because of the often inconsistent, incomplete and sometimes unknown uncertainties involved in the analyses. Thus, quantifying uncertainties in extracting the \es0 and \l0 from model analyses of isovector observables is urgently needed.
Indeed, such efforts are currently being made by several groups, see, e.g., refs. \cite{ste14,JPG}.

\section{WHY IS THE SYMMETRY ENERGY SO UNCERTAIN ESPECIALLY AT SUPRA-SATURATION DENSITIES}
\vspace{-0.5cm}
\begin{figure}[htb]
\includegraphics[width=12.cm,height=4.2cm]{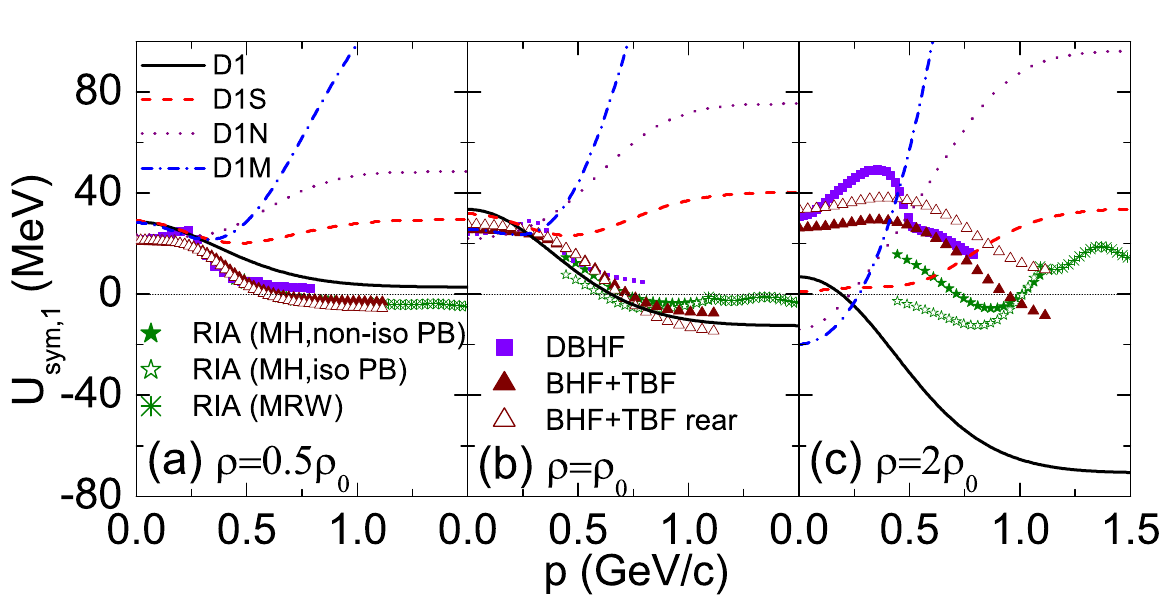}
\caption{Density and momentum dependence of the nucleon isovector potential predicted by the Gogny-Hartree-Fock calculations using the D1, D1S, D1M and D1N interactions, Dirac-Brueckner-Hartree-Fock (DBHF) and Relativistic Impulse Approximation (RIA) with various two-body and three-body forces (TBF) \protect\cite{Rchen}.}\label{Usym1}
\end{figure}
Why is the \esym so uncertain especially at high densities? This is a hard question to answer. Moreover, the attempted answers are most likely to be model dependent. In this section, we first try to get some general hints at the mean-field level using the Hugenholtz-Van Hove (HVH) theorem \cite{hug}. We will then illustrate effects of the three-body and tensor force on the \esym. It is well known that the single-particle potential $U_{n/p}(\rho,k)$ for nucleons with momentum $k$ in nuclear matter of density $\rho$ and isospin asymmetry $\delta$ can be well approximated by the so-called Lane form \cite{Lan62}
$
U_{n/p}(\rho,k)\approx U_0(\rho,k) \pm U_{sym,1}(\rho,k)\cdot\delta 
$
in terms of the nucleon isoscalar potential $U_0(\rho,k)$ and isovector (symmetry) potential $\pm U_{sym,1}(\rho,k)$. The $\pm$ sign indicates that the symmetry potential is repulsive for neutrons and attractive for protons in neutron-rich matter consistent with findings from optical model analyses of nucleon-nucleus scattering data since the 1960's. Using the Brueckner theory \cite{bru64,Dab73} or more generally the Hugenholtz-Van Hove (HVH) theorem \cite{hug}, the \esym and its slope $L(\rho)$
at an arbitrary density $\rho$ can be expressed as 
\begin{equation}
E_{\rm sym}(\rho) = \frac{1}{3} \frac{\hbar^2 k_F^2}{2 m_0^*} +
\frac{1}{2} U_{\rm sym,1}(\rho,k_{F})~~
{\rm and}~~ 
L(\rho) = \frac{2}{3} \frac{\hbar^2 k_F^2}{2 m_0^*} +
\frac{3}{2} U_{\rm sym,1}(\rho,k_F) + \frac{\partial U_{\rm sym,1}}{\partial k}|_{k_F} k_F
\end{equation}
where $k_F$ is the nucleon Fermi momentum and $m^*_0$ is the nucleon isoscalar effective mass \cite{XuLiChen10a,xuli2,Rchen}.
While the density and momentum dependence of the isoscalar potential $U_{0}(\rho,k)$ has been relatively well determined \cite{dan02}, such information for the isovector potential $U_{sym,1}(\rho,k)$ is rather incomplete especially at high-densities and/or high-momenta \cite{LCK08,Fuchs05,Fuchs06,zuo05,ria05,ria06}. As examples, shown in Fig.\ \ref{Usym1} are the predicted isovector potentials using the Gogny-Hartree-Fock, Dirac-Brueckner-Hartree-Fock and Relativistic Impulse Approximation with various two-body and three-body  interactions \cite{Rchen}. It is seen that while some models predict decreasing symmetry potentials albeit at different rates,
some others predict instead increasing ones with growing nucleon momentum especially at high densities. For instance, the four widely used Gogny interactions \cite{Gogny} D1, D1S, D1N and D1M predict very different high-momentum behaviors for the isovector potential. Thus, the high-density and/or -mometum behavior of the isovector potential $U_{sym,1}(\rho,k)$ is the key physics ingredient governing the symmetry energy and properties of dense neutron-rich nucleonic matter.  Of course,  different approaches used in treating quantum many-body problems also contribute to the divergence of the predicted symmetry energy especially at supra-saturation densities.

Going one step further, one may ask why the isovector potential is so uncertain especially at high-densities and/or high-momenta. To answer this question, it is instructive to look at the expression of the isovector potential at $k_F$ in the interacting Fermi gas model \cite{pre,Xu-tensor} 
$
U_{sym,1}(k_F,\rho)=
\frac{1}{4}\rho\int [V_{T1}(r_{ij})f^{T1}(r_{ij})-V_{T0}(r_{ij})f^{T0}(r_{ij})]d^3r_{ij}
$
in terms of the isosinglet (T=0) and isotriplet (T=1)
nucleon-nucleon (NN) interactions $V_{T0}(r_{ij})$ and
$V_{T1}(r_{ij})$, and the corresponding NN correlation functions
$f^{T0}(r_{ij})$ and $f^{T1}(r_{ij})$, respectively. Needless to
say, if there is no isospin dependence in both the NN interaction
and the correlation function, then the isovector potential $
U_{sym,1}(k_F,\rho)$ vanishes. The $E_{sym}(\rho)$ thus reflects the
competition of the NN interaction strengths and correlation
functions between the isosinglet and isotriplet channels.
Among the key factors affecting the competition are (1) the spin-isospin dependence of the three-body force, (2) tensor forces mostly in the isosinglet channel and (3) the isospin dependence of nucleon-nucleon correlations \cite{Xu-tensor}. Our poor knowledge about the in-medium properties of these factors, such as the uncertain short-range behavior of the tensor force due to $\rho$-meson exchange, contribute dominantly to the uncertain density and momentum dependence of the isovector potential especially at supra-saturation densities \cite{Xu-tensor,Xulili,Wang12}.
\vspace{-0.28cm}
\begin{figure}[htb]
\includegraphics[width=5.6cm,height=4.cm]{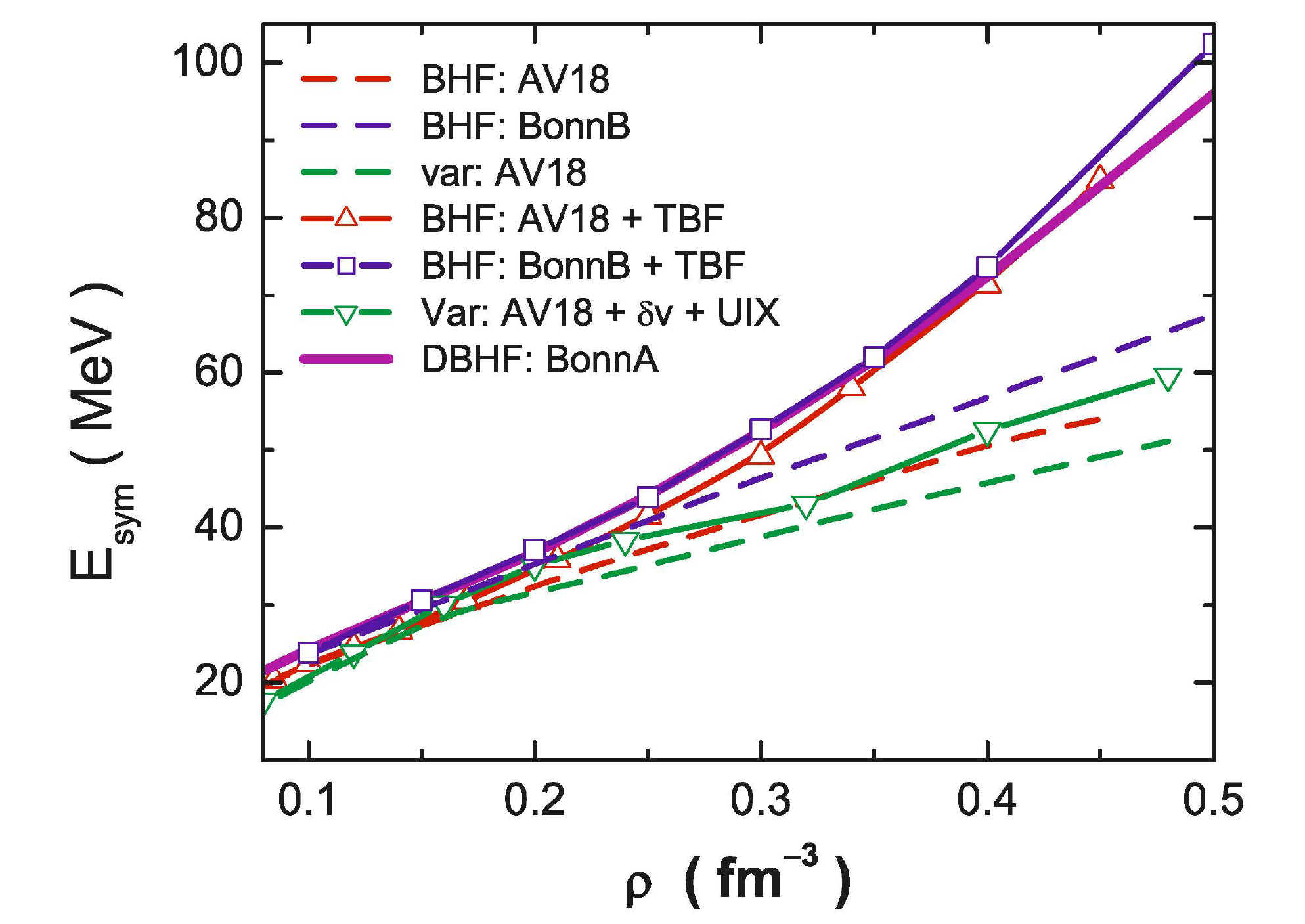}
\includegraphics[width=5cm,height=4cm]{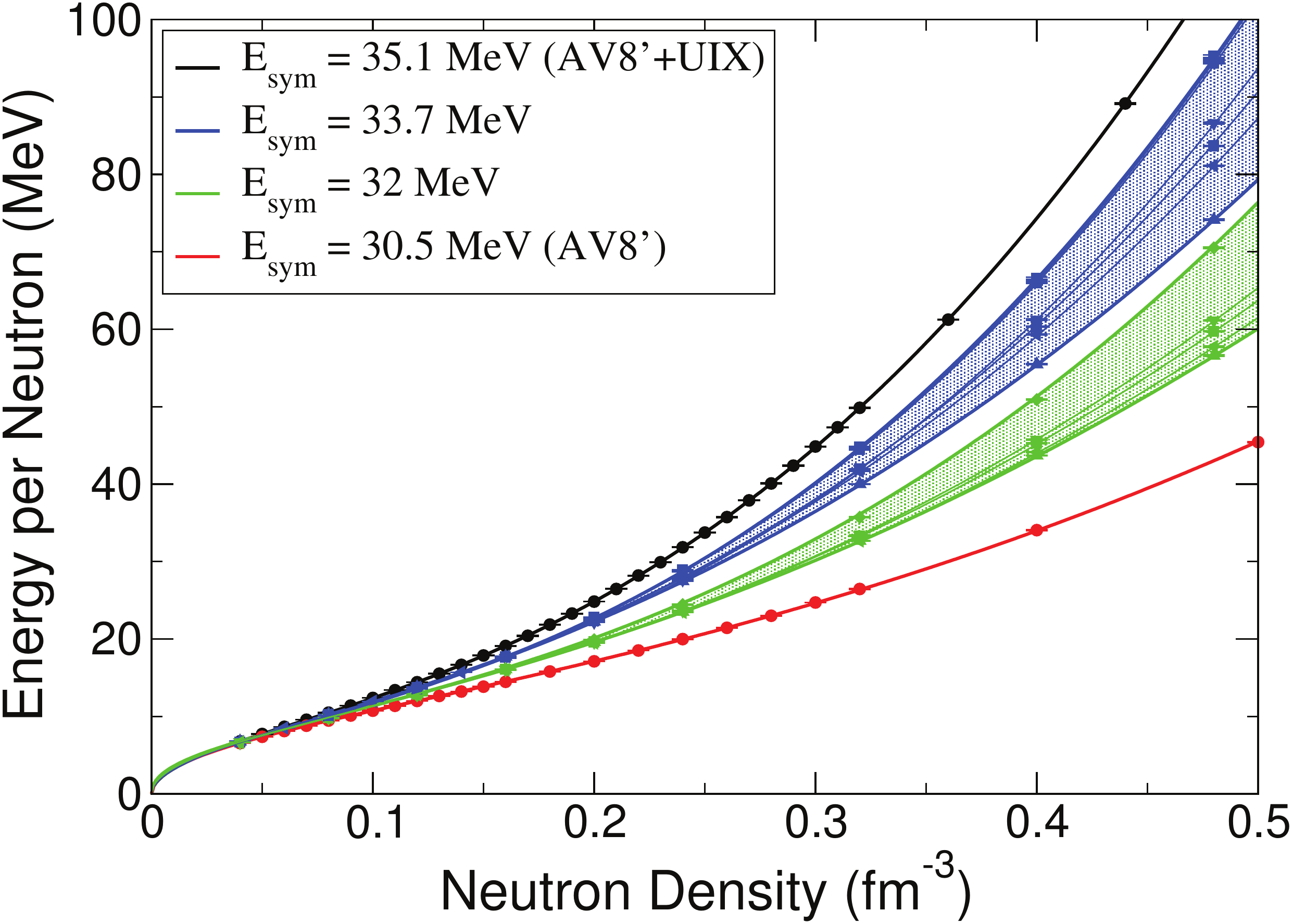}
\includegraphics[width=5.5cm,height=4.3cm]{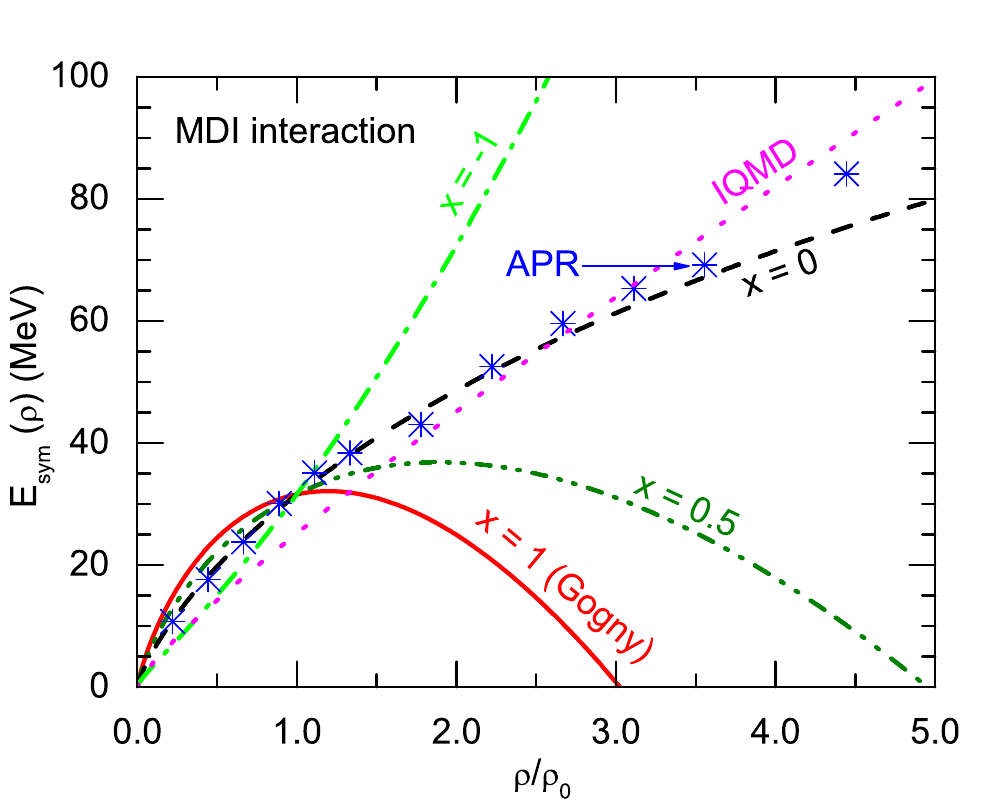}
\caption{{\protect Left: \esym from three different ab initio theoretical frameworks with and without the three-body forces \cite{Zuo14}.
Middle: The equation of state of pure neutron matter obtained by using various models of three-neutron forces for a given symmetry energy at $\rho_0$ \cite{Gan14}.\label{Gan}
Right: \esym with different strengths of three-body force within the Hartree-Fock approach using the modified Gogny interaction (MDI with different x parameters) in comparison with the APR prediction \cite{XiaoPRL}. }}
\end{figure}
\vspace{-0.3cm}
\subsection{Three-body force and high-density symmetry energy}
Effects of the three-body forces on the density dependence of nuclear symmetry energy is well known, see, e.g., ref.\ \cite{Zuo14} for a recent review.  As an example, shown in the left window of Fig.\ \ref{Gan} are the \esym from three different theoretical frameworks with and without the three-body forces: the BHF and Variational Many-Body approach (non-relativistic) and the DBHF (relativistic). It is well known that the three-body force is necessary for the non-relativistic models to reproduce properly the saturation properties of symmetric nuclear matter (SNM). It is seen that the repulsive three-body force stiffens the symmetry energy at supra-saturation densities. It is interesting to note that the BHF with either the AV18+TBF or the BonnB+TBF interactions can reproduce the prediction of the DBHF with the BonnA interaction, indicating that both the two-body and three-body forces are at work. There are obviously degeneracies/ambiguities in using different combinations of the two-body and three-body interactions.
Similar conclusions were also found in studying the EOS of pure neutron matter (PNM). As an example, shown in the middle of Fig.\ \ref{Gan} are the AFQMC predictions for the PNM EOS with and without three-body forces \cite{Gan14}. Comparing the red and black lines, it is seen that with the UIX three-body force, while the symmetry energy at $\rho_0$ only increases a little the PNM EOS stiffens a lot at supra-saturation densities. The empirical UIX three-body force has four parts describing different three-body interaction mechanisms \cite{Gan14}. By modifying some properties of these four parts, such as the strength and range parameter of the short-range terms, one can modify the PNM EOS at supra-saturation densities while keeping the symmetry energy at $\rho_0$ a constant. The blue lines (all with $E_{sym}(\rho_0)=33.7$ MeV) and green lines (all  with $E_{sym}(\rho_0)=32$ MeV) are examples of such studies.  Since the $E_{sym}(\rho_0)$ is mainly determined by the two-body forces at $\rho_0$, the comparison between the blue and green bands illustrates the interplay of the two-body and three-body forces in determining the PNM EOS. Thus, properties of the PNM EOS, and subsequently the symmetry energy which is the difference between the EOS of PNM and SNM, at supra-saturation densities depend sensitively on properties of both the two-body and three-body forces. 

In many phenomenological models for studying both nuclear structures and heavy-ion reactions, 
see, e.g., Refs. \cite{vau,Oni78,dec,Gra89}, one often uses a
zero-range three-body force. It is normally represented by a density-dependent effective two-body force after integrating over the third nucleon, i.e.,
$
V_{d}=t_0(1+x_0P_{\sigma})\rho^{\alpha}\delta(r),
$
where $t_0$, $\alpha$ and $x_0$ are parameters and $P_{\sigma}$ is
the spin-exchange operator.  Its contribution to the symmetry energy is 
$
E_{sym}^{TBF}= -(1+2x_0)\frac{t_0}{8}\rho ^{\alpha +1}.
$
The parameter $x_0$ controling the spin-isospin dependence of the three-body force also determines its contribution to the symmetry energy $E_{sym}^{TBF}$. 
In the original Gogny force \cite{dec} with $x_0=1$ and $\alpha=1/3$, the symmetry
energy $E_{sym}(\rho)$ drops quickly to zero at about $3\rho_0$. The same three-body force with varying $x_0$ has been
used in the Momentum Dependent Interaction (MDI) \cite{Das03} often used in transport model simulations of heavy-ion collisions \cite{IBUU04,Cozma}. 
Shown in the right window of Fig. \ \ref{Gan} is the $E_{sym}(\rho)$  with the MDI interaction using several values of $x\equiv (1+2x_0)/3$ to mimic divergent predictions of high-density symmetry energy. It is seen that by varying the spin-isospin strength $x$ of the three-body force, thus the competition of the isosinglet and isotriplet interactions, the $E_{sym}(\rho)$ goes from negative to very positive at high densities \cite{XiaoPRL}.
\vspace{-0.3cm}
\subsection{Tensor force and high-density symmetry energy}
It is well known that the nuclear force due to the pion ($\rho$) meson exchange has an intermediate (short) range attractive (repulsive) tensor component. It has been shown that the tensor force influences significantly the  $E_{sym}(\rho)$ in many studies \cite{kuo65,bro81,sob,Pan72,Wir88a,bro1,mac94,Eng98,xuli,Lee1,Lee2,Car13} using various approaches ranging from simple phenomenological models to state-of-the-art microscopic many-body theories. Effects of the tensor force on the high-density behavior of $E_{sym}(\rho)$ vary broadly in ways very similarly to varying the form/strength of the spin-isospin dependence of the three-body force. Ironically, in exactly the same way as for reproducing the empirical properties of nuclear matter at saturation density, one can adjust the strength of either the three-body force or tensor force to give the same \esym. This was demonstrated clearly in refs. \cite{xuli,Dong} using both phenomenological and microscopic models. This makes it even harder to trace the origins of the uncertain \esym as many models do not consider the tensor force/coupling at all but have other mechanisms to describe properties of both finite nuclei and nuclear matter equally well. In most transport models for heavy-ion reactions, as tensor force has no contribution to the mean-field for spin-saturated systems, effects of the tensor force has not been considered directly. However, we notice that efforts are being made in further developing spin-isospin dependent transport models for studying effects of the spin-orbit coupling and/or tensor force on the spin and isospin transport in nuclear reactions \cite{Xu-spin,Jun15}. 
\vspace{-0.45cm}
\begin{figure}[h]
\begin{minipage}{12.2pc}
\includegraphics[scale=0.23]{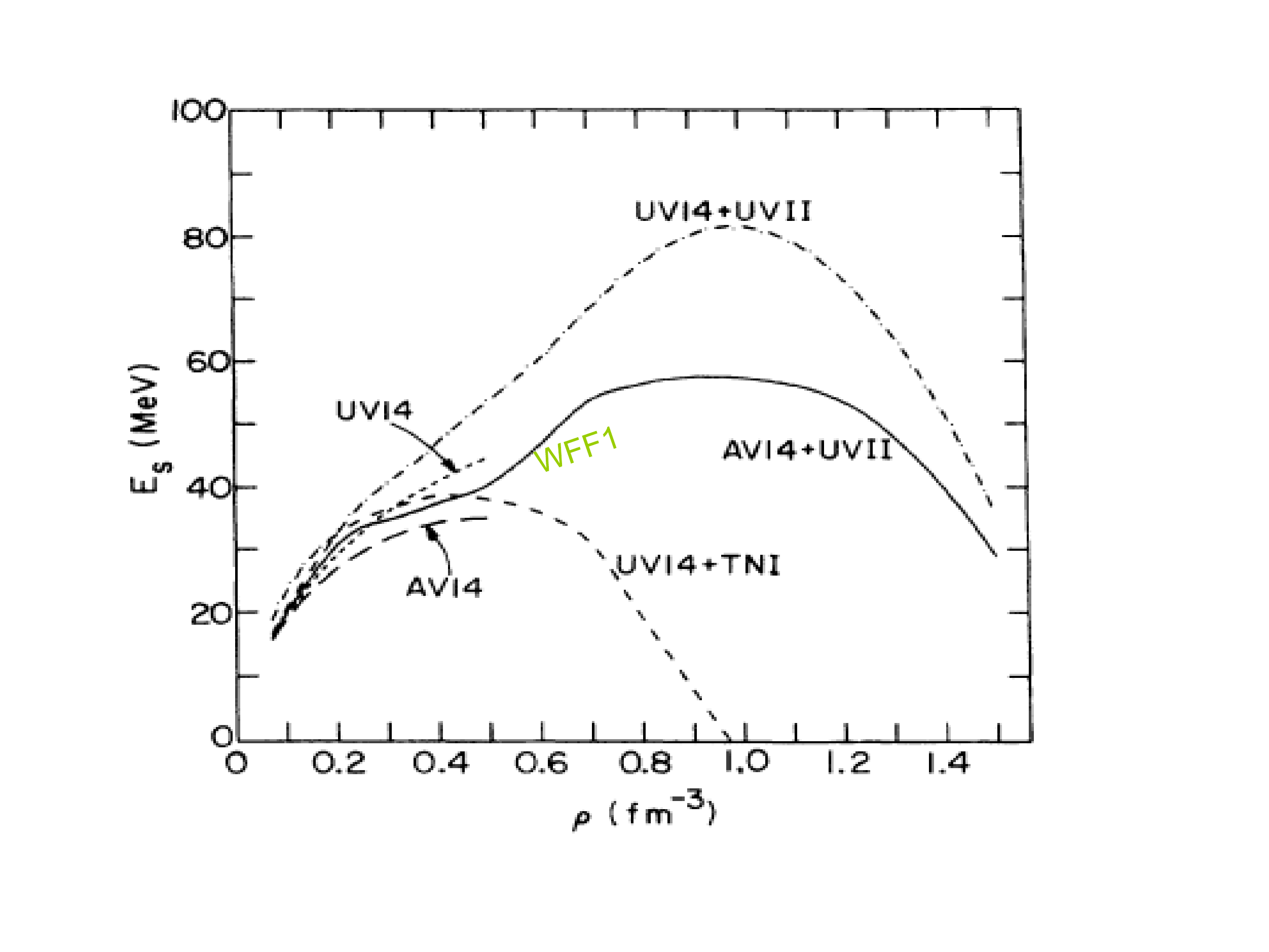}
\end{minipage}
\begin{minipage}{12.3pc}
\includegraphics[scale=0.26]{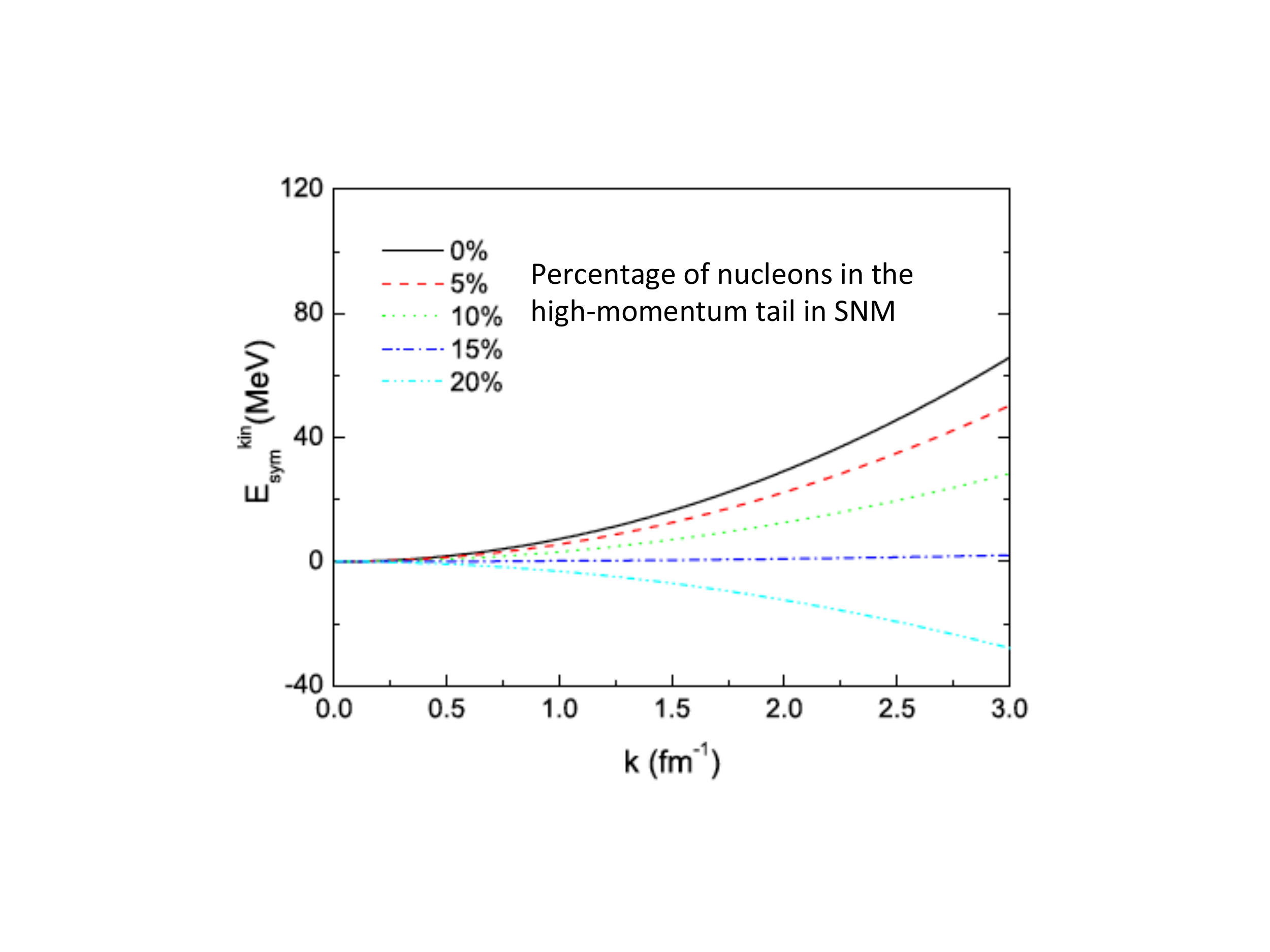}
\end{minipage}
\begin{minipage}{12.2pc}
\includegraphics[scale=0.26]{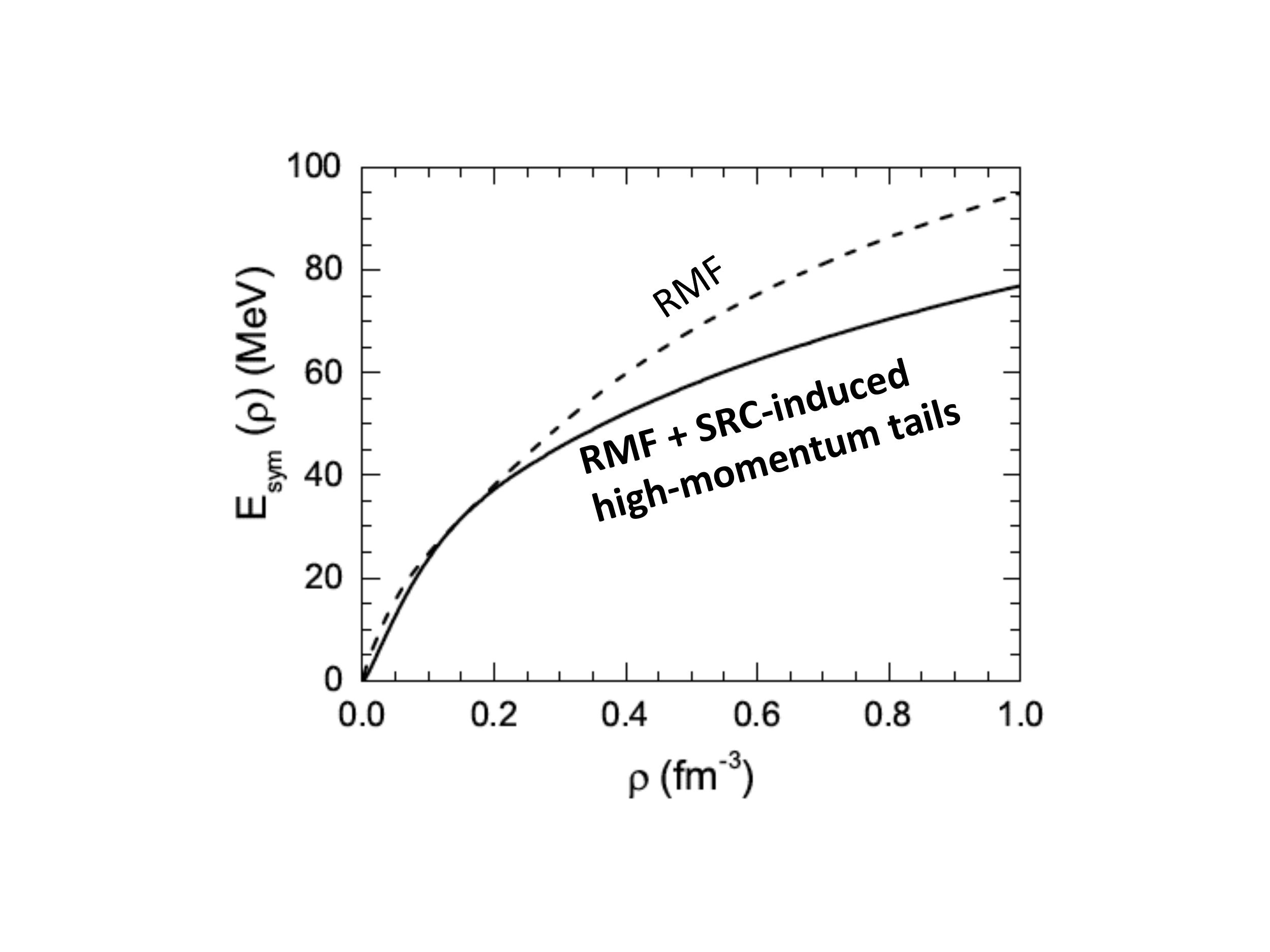}
\end{minipage}
\vspace{-10cm}
\caption{{\protect Left: \esym from the Variational Many-Body Theory using different two- and three-body interactions by Wiringa et al. \cite{Wir88a}.
Middle: The kinetic symmetry energy of correlated nucleons as a function of Fermi momentum with different fractions of high-momentum nucleons in SNM \cite{Xulili}. 
Right: \esym within the traditional RMF (dashed) or RMF incorporating the SRC-modified single-nucleon momentum distribution with a high-momentum tail \cite{Cai-RMF}.}}\label{SRC}
\end{figure}

Two interesting and direct effects of the tensor force on the high-density $E_{sym}(\rho)$ deserve special discussions and they might be probed indirectly using heavy-ion collisions or properties of neutron stars. Firstly, it is well known that the tensor force acts mostly in the isosinglet n-p channel. When the repulsive tensor force due to the $\rho$ meson exchange dominates at high densities, the potential energy in SNM can increase faster than that in PNM where the tensor force is negligible, leading to a negative symmetry energy above certain densities \cite{Pan72,Wir88a}. Shown in the left window of Fig.\ \ref{SRC} is an example of variational many-body predictions on effects of the three-body force and/or tensor force at high densities. A summary of more predictions of similar high-density behavior of symmetry energy can be found in Ref. \cite{Kut06}. 
The decreasing/negative symmetry energy at high densities leads to the interesting possibility of forming proton polarons \cite{Kut93,Kut94} in neutron-rich nucleonic matter, the need for a modified gravity in massive neutron stars \cite{Wen09,Wlin14,Jiang15} or the existence of a weakly interacting light boson mediating a new force \cite{Yong13,Kra16,Feng16}. 

Another interesting effect of the tensor force is in reducing the kinetic symmetry energy \cite{Xulili}. It is well known that the SRC induced by the tensor force mostly for n-p pairs leads to a high-momentum tail in the single-nucleon momentum distribution above the Fermi surface especially in SNM \cite{bethe,Pan97,Cio15}. Based on the information extracted from experiments done at the Brookhaven and Jefferson National Laboratories \cite{Sub08,Arr12,Hen14}, the percentages of nucleons in the high momentum tail is estimated to be about 25\% in SNM and 1-2 \% in PNM. Theoretical calculations predict about 10-25\% high momentum nucleons for SNM and 1-5\% for PNM depending on the model and interaction used \cite{VMC,mu04,Rio09,Rio14,Yin13,ZHLi}. A direct consequence of the isospin dependence of the SRC is that the nucleon average kinetic energy is increased by the SRC more in SNM than PNM. Since the symmetry energy is essentially the difference in energy between the PNM and SNM within the parabolic approximation, one expects the SRC to reduce the kinetic symmetry energy with respect to the value of $E_{\rm sym}^{\rm Kin}(\rm{FFG})=12.5(\rho/\rho_0)^{2/3}$ one normally uses in textbooks for an uncorrelated free Fermi gas (FFG). For example, shown in the middle of  Fig.\ref{SRC} is the kinetic symmetry energy as a function of Fermi momentum with different fractions of high-momentum nucleons in SNM while that in PNM is set to be zero. It is interesting to see that the tensor force induced SRC has a significant impact on the kinetic symmetry energy, especially at supra-saturation densities. With about 15\% high momentum nucleons in SNM, the $E_{sym}^{kin}(\rho)$ is almost zero in a broad range of Fermi momentum. With more nucleons in the high momentum tail, the kinetic symmetry energy becomes negative at higher densities. This expectation first made based on a phenomenological model \cite{Xulili,Xu11} has been confirmed at various quantitative levels in studies based on the Brueckner-Hartree-Fock approach (BHF) \cite{Vid11}, the Self-Consistent Green's Functions approach (SCGF) \cite{Car12}, and the Fermi-Hypernetted-Chain calculations (FHNC) \cite{Lov11}. Moreover, as shown in the right window of Fig. \ref{SRC}, incorporating the high-momentum tail in the kinetic part of the RMF energy density while reproducing the same empirical properties of SNM as well as the symmetry energy and its slope $L$ at saturation density as the original RMF, the \esym becomes more concave. Consequently, the experimentally measured curvature of the symmetry energy (or the isospin dependence of the incompressibility $K_{\tau}$) can be better reproduced \cite{Cai-RMF}.  

Generally, in energy density functional theories one writes the total energy as the sum of a potential energy and a kinetic energy of free particles having a step function for their momentum distribution at zero temperature. 
Model parameters in the potential energy are then determined by minimizing the total energy and reproducing all empirical properties of nuclear matter as well as sometimes some properties of selected nuclei. Considering the SRC induced by the tensor force, the single-nucleon momentum distribution is modified from the step function and one is now dealing with quasi-particles instead of free nucleons. The kinetic symmetry energy of these quasi-particles is smaller than the $E_{\rm sym}^{\rm Kin}(\rm{FFG})$ that is only due to the Pauli exclusion principle and the different Fermi surfaces for neutrons and protons. In many studies in both nuclear physics and astrophysics, it is customary to write the total symmetry energy as $E_{sym}(\rho)=12.5(\rho/\rho_0)^{2/3}+E^{pot}_{sym}(\rho)$. Often the potential symmetry energy is parameterized as $E^{pot}_{sym}(\rho)=[E_{sym}(\rho_0)-12.5](\rho/\rho_0)^{\gamma}$ to conserve the total \esym at $\rho_0$. The exponential $\gamma$ is then determined by fitting some experimental data or observations.  In doing so, one puts all interaction effects on the potential symmetry energy and the kinetic part is assumed to be the same as for free nucleons.  In most astrophysical applications of the EOS, such as solving the TOV equation, one use the relationship between the pressure and energy density. It does not matter how the energy density/pressure is split between its kinetic and potential parts as only its total goes into the TOV equation as recently shown in Ref. \cite{Hen16Steiner}. However, in calculating the critical densities for forming different charge states of the $\Delta(1232)$ resonances in neutron stars using chemical equilibrium conditions, it appears that how the \esym is split between its kinetic and potential parts is important as they have different density dependences \cite{Cai-delta}. Moreover, in simulating heavy-ion collisions based on Boltzmann-type transport equations for quasi-particles, how the \esym is split into kinetic and potential parts does matter and the reaction dynamics is different \cite{LiG15,Hen15b}. For example, with the reduced kinetic contribution, the potential symmetry energy has to be increased to keep the same total \esym. Then, the corresponding symmetry potential has to be enhanced. This will lead to enhanced dynamical effects for isospin tracers \cite{LiG15,Hen15b}. Furthermore, using the Migdal-Luttinger theorem\,\cite{Mig57,Lut60}, one can directly relate the size of the SRC with the E-effective masses of nucleons, namely the lifetime of the quasi-particles \cite{Cai-Emass}. In-medium nucleon-nucleon total and differential cross sections are also inputs to transport models. However, at this point it is not clear to us how the tensor-force induced SRC affects these cross sections, and whether/how multi-nucleon scatterings involving off-shell nucleons should be considered. Thus, there are still many interesting questions regarding the tensor force effects on high-density \esym and effective ways to probe them using heavy-ion reactions. 

\section{ONGOING EFFORTS TO CONSTRAIN HIGH-DENSITY SYMMETRY ENERGY}
How can nuclear reactions in terrestrial laboratories help constrain the high-density behavior of \esym? How do we probe the underlying nucleon isovector potential, its momentum dependence and the corresponding neutron-proton effective mass splitting? What are the effects of clustering and pairing on the low-density symmetry energy? What are the information content of the isovector reaction observables to be measured with advanced detectors currently under construction? How to quantify theoretical uncertainties of transport models used to extract nuclear symmetry energy from reaction observables?  These are among the key questions the low-intermediate energy heavy-ion reaction community has been trying to address in recent years, see, e.g., refs. \cite{ireview98,ibook01,dan02,ditoro,LCK08,Lynch09,Trau12,Tsang12,Chuck14,Sherry14,Fil14,Ade14,Khoa,Joe14,Alan,Hud14}. While significant progresses have been made, much more work remains to be done. Heavy-ion collisions provide a unique mechanism to produce dense neutron-rich nucleonic matter in terrestrial laboratories.
\vspace{-0.35cm}
\begin{figure}[h]
\begin{minipage}{14.5pc}
\includegraphics[scale=0.29]{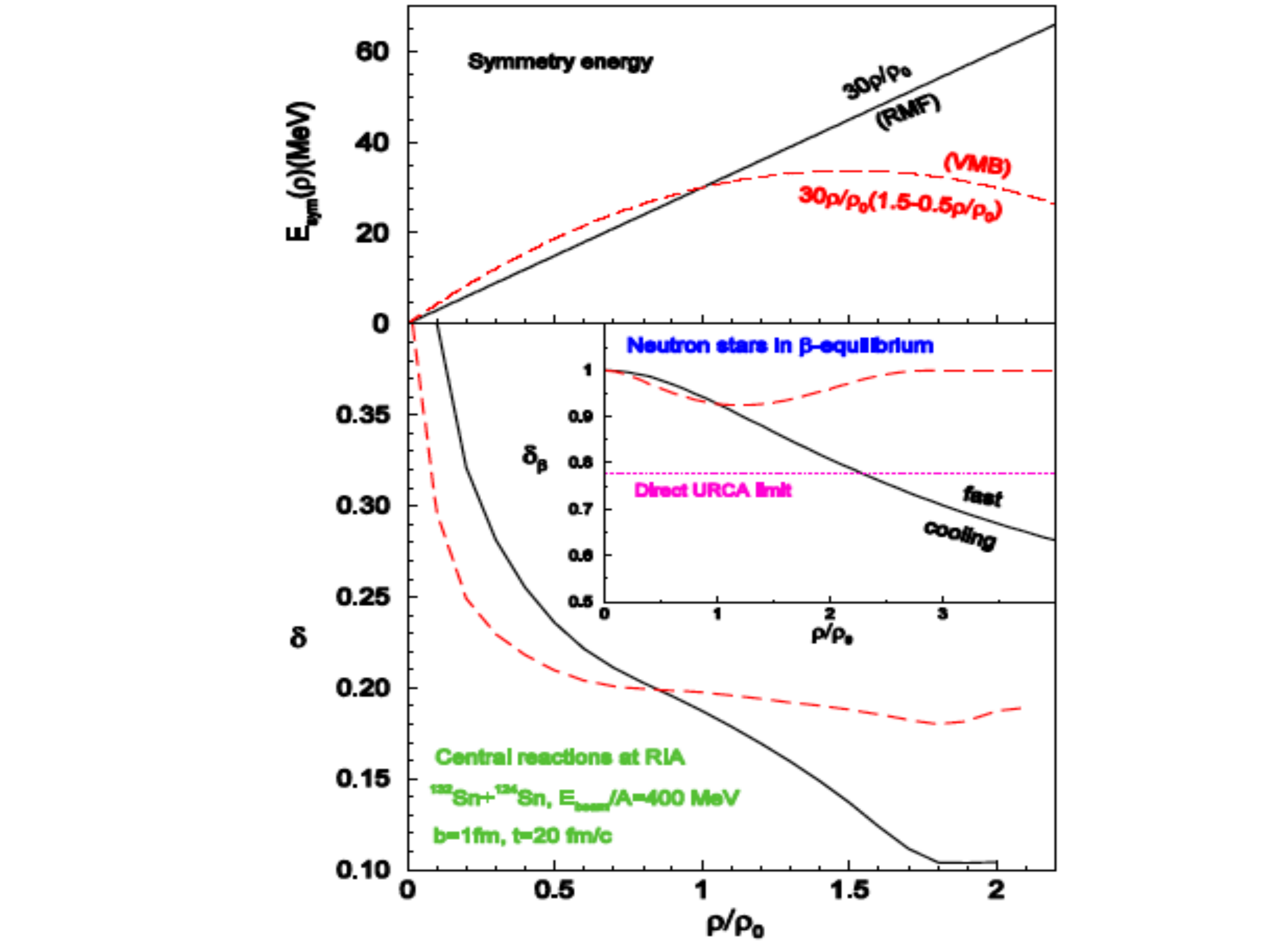}
\end{minipage}
\begin{minipage}{14.5pc}
\includegraphics[scale=0.32]{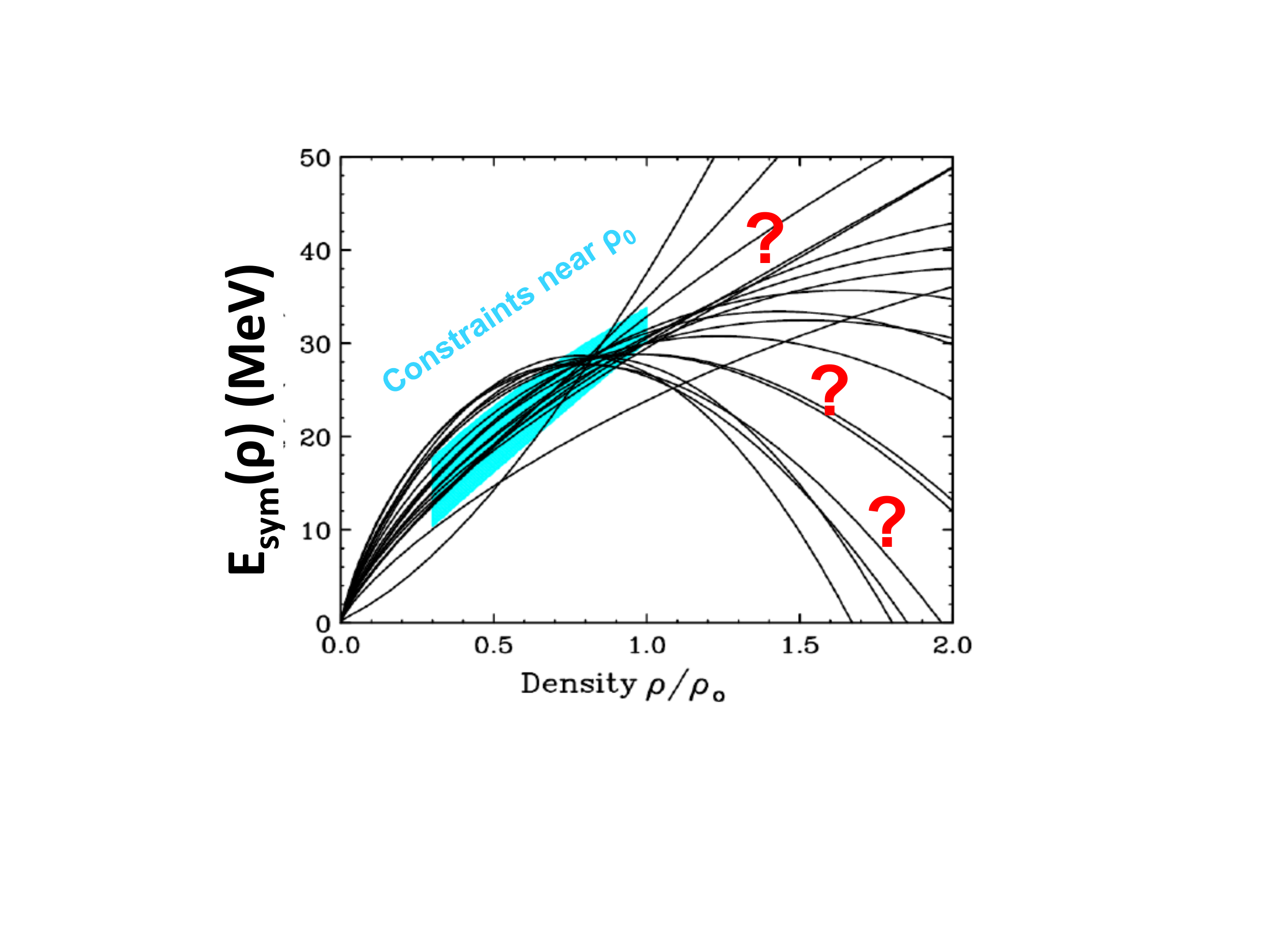}
\end{minipage}
\begin{minipage}{14.5pc}
\includegraphics[scale=0.27]{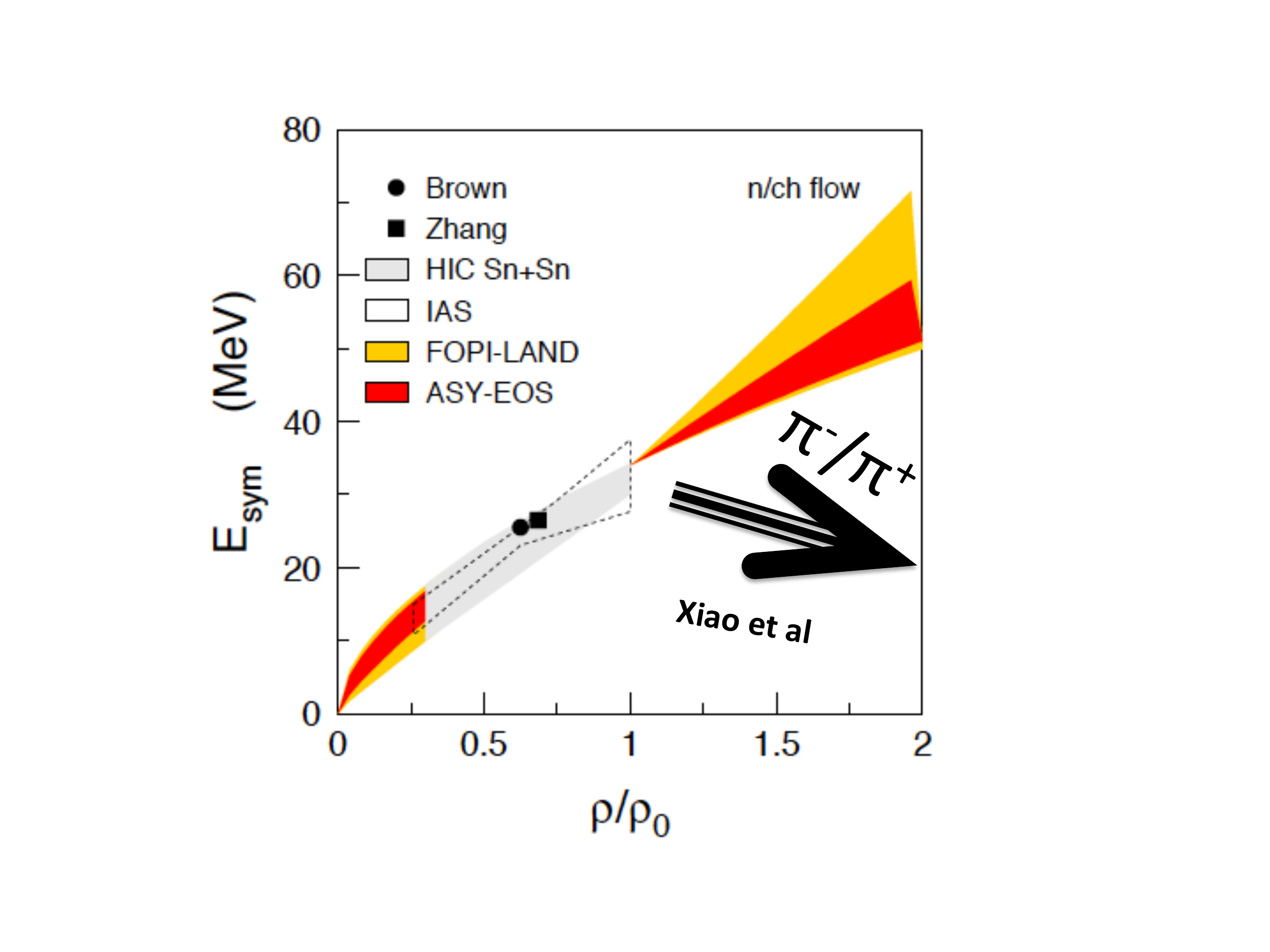}
\end{minipage}
\caption{{\protect Left: The isospin asymmetry-density correlation at t=20 fm/c over the whole space in the central $^{132}$Sn +$^{124}$Sn reaction at a beam energy of 400 MeV/nucleon with the two different forms of \esym shown in the upper box, respectively. The corresponding correlation in neutron stars at $\beta$ equilibrium is shown in the inset \cite{Li-PRL}.
Middle: The shaded area covers the extracted constraints on the \esym near saturation density from isospin diffusion experiments \cite{Tsang12} in comparison with Skyrme Hartree-Fock calculations \cite{brown,Stone}. Right: \esym extracted by the ASY-EOS Collaboration \cite{ASY-EOS} using UrQMD/QMD in comparison with the trend (arrow) of \esym from an earlier analysis of the FOPI pion data using IBUU04 by Xiao et al. \cite{XiaoPRL} between $\rho_0$ and $2\rho_0$ as well as other constraints at lower densities.}}\label{Exp}
\end{figure} 
As an illustration, shown in the left window of Fig. \ref{Exp} is the correlation between the isospin asymmetry $\delta$ and density at the instant of 20 fm/c in central reactions of $^{132}$Sn+$^{124}$Sn at a beam energy of 400 MeV/A \cite{Li-PRL}. It is clearly seen that the isospin asymmetry of dense nucleonic matter at about $2\rho_0$ reached during the reaction is very sensitive to the high-density behavior of the \esym shown in the upper frame. 
The inset is the corresponding correlation in neutron stars at $\beta$ equilibrium. Interestingly, the isospin asymmetry-density correlations in heavy-ion collisions and neutron stars are very similar.  
Isospin tracers and observables are mostly relative quantities of neutron-proton pairs or mirror nuclei as the isovector potential at isospin asymmetries reached in heavy-ion reactions is very small compared to the isosclar potential. One can minimize/maximize effects of the isoscalar/isovector potential by using these relative observables. Moreover, often one has to construct isospin-sensitive observables by using several reactions to cancel out isoscalar or uncertainties mainly due to the difficulties of accurately measuring neutrons. For example, shown in the middle window is the constraint on the symmetry energy near the saturation density from isospin diffusion experiments using 4 reactions involving $^{112}$Sn and $^{124}$Sn \cite{Tsang-PRL}. Currently, major efforts are being made by the SEP (Symmetry Energy Project) Collaboration \cite{SEP} and the ASY-EOS Collaboration \cite{ASY-EOS} to constrain the high-density \esym. While the SEP experiments currently focus on measuring the $\pi^-/\pi^+$ ratio, the ASY-EOS Collaboration has recently analyzed the relative flows of neutrons w.r.t. protons, tritons w.r.t. $^3$He and yield ratios of light isobars. Shown in the right window of Fig.\ \ref{Exp} are the \esym they extracted using two versions of transport models based on the Quantum Molecular Dynamics in comparison with an earlier result from analyzing the $\pi^-/\pi^+$ data from the FOPI collaboration using a BUU-type transport model \cite{XiaoPRL}. Obviously, there is a disagreement about the trend of the \esym at supra-saturation densities. Trusting the data, whether the \esym is stiff or super-soft at supra-saturation density remains an open issue \cite{XiaoPRL,russ11,Feng10,cozma11}. Realizing the importance of understanding the model dependence, serious efforts are being made by the transport code developers and users to better understand very often different techniques used in modeling various processes happening during heavy-ion reactions \cite{Xu-code}.

It is well known that the radii of neutron stars are most sensitive to the \esym around $\rho_0-2\rho_0$  \cite{Steiner05,Lat13}. Thus, a precise measurement of neutron star radii is another way of constraining the high-density behavior of \esym. However, there are several longstanding difficulties in measuring accurately the radii of neutron stars. Reviews on recent progress and new challenges in extracting the high-density \esym from astrophysical observations of neutron stars and gravitational waves can be found in, e.g., Refs. \cite{ste14,Newton14,Far-13b,Plamen16}.

\section{SUMMARY AND ACKNOWLEDGEMENT}
The high-density behavior of \esym is the most uncertain part of the EOS of neutron-rich nucleonic matter. It depends strongly on the poorly known spin-isospin dependence of the three-body force, the
short-range behavior of nuclear tensor force and the resulting isospin-dependent short-range correlations. The high-density behavior of \esym has many interesting ramifications for heavy-ion reactions, properties of neutron stars and gravitational waves as well as the nature of strong-field gravity. Significant progresses have been made in better understanding the underlying physics determining the high-density \esym thanks to the hard work of many people in both nuclear physics and astrophysics communities. Dedicated heavy-ion experiments are being carried out to probe the high-density \esym in terrestrial laboratories. Coordinated efforts in developing more reliable theoretical tools for extracting the high-density \esym from the new experiments are underway. Various efforts of extracting the high-density \esym using observations of neutron stars and gravitational waves are also underway. 

This lecture is mainly based on the results of my collaborations with Bao-Jun Cai, Lie-Wen Chen, Farrooh Fattoyev, Wenjun Guo, Xiao-Tao He, Or Hen, Plamen Krastev, Wei-Zhou Jiang, Che Ming Ko, Ang Li, Xiao-Hua Li, Eli Piasetzky, William G. Newton, Zhaozhong Shi, Andrew Steiner, De-Hua Wen, Larry B. Weinstein, Chang Xu, Jun Xu, Gao-Chan Yong and Wei Zuo. This work is supported in part by the U.S. Department of Energy, Office of Science, under Award Number DE-SC0013702 and the National Natural Science Foundation of China under Grant No. 11320101004.


\end{document}